\documentclass[aps,tightenlines,twocolumn]{revtex4-1}
\usepackage{amsmath,amsfonts,amssymb,graphicx}
\usepackage{times}
\usepackage{dcolumn}
\usepackage{bm}
\usepackage{lineno}

\makeatletter
\renewcommand{\fnum@figure}{FIG. \thefigure}
\renewcommand{\fnum@table}{TABLE \thetable}
\makeatother

\newcommand{\ems}{\sqrt{s_{NN}}}
\newcommand{\vone}{v_{1}(y)}
\newcommand{\dvdy}{dv_{1}/dy}

\def\Journal#1#2#3#4{{#1} {\bf #2}, #3 (#4)}

\def\NIMA{Nucl. Instrum. Methods A}
\def\NPA{Nucl. Phys. A}
\def\PRL{Phys. Rev. Lett.}
\def\PRC{Phys. Rev. C}

\def\PLB{Phys. Lett. B}

\begin{document}

\title{\quad\\[0.5cm]
\boldmath Beam-Energy Dependence of the Directed Flow of Deuterons in Au+Au Collisions}
\date{\today}

\author{
J.~Adam$^{6}$,
L.~Adamczyk$^{2}$,
J.~R.~Adams$^{39}$,
J.~K.~Adkins$^{30}$,
G.~Agakishiev$^{28}$,
M.~M.~Aggarwal$^{41}$,
Z.~Ahammed$^{61}$,
I.~Alekseev$^{3,35}$,
D.~M.~Anderson$^{55}$,
A.~Aparin$^{28}$
% ...
}

\collaboration{STAR Collaboration}

\address{$^{1}$Abilene Christian University, Abilene, Texas   79699}
\address{$^{2}$AGH University of Science and Technology, FPACS, Cracow 30-059, Poland}
\address{$^{3}$Alikhanov Institute for Theoretical and Experimental Physics NRC "Kurchatov Institute", Moscow 117218, Russia}
\address{$^{4}$Argonne National Laboratory, Argonne, Illinois 60439}
\address{$^{5}$American University of Cairo, New Cairo 11835, New Cairo, Egypt}
\address{$^{6}$Brookhaven National Laboratory, Upton, New York 11973}
\address{$^{7}$University of California, Berkeley, California 94720}
\address{$^{8}$University of California, Davis, California 95616}
\address{$^{9}$University of California, Los Angeles, California 90095}
\address{$^{10}$University of California, Riverside, California 92521}
\address{$^{11}$Central China Normal University, Wuhan, Hubei 430079 }
\address{$^{12}$University of Illinois at Chicago, Chicago, Illinois 60607}
\address{$^{13}$Creighton University, Omaha, Nebraska 68178}
\address{$^{14}$Czech Technical University in Prague, FNSPE, Prague 115 19, Czech Republic}
\address{$^{15}$Technische Universit\"at Darmstadt, Darmstadt 64289, Germany}
\address{$^{16}$ELTE E\"otv\"os Lor\'and University, Budapest, Hungary H-1117}
\address{$^{17}$Frankfurt Institute for Advanced Studies FIAS, Frankfurt 60438, Germany}
\address{$^{18}$Fudan University, Shanghai, 200433 }
\address{$^{19}$University of Heidelberg, Heidelberg 69120, Germany }
\address{$^{20}$University of Houston, Houston, Texas 77204}
\address{$^{21}$Huzhou University, Huzhou, Zhejiang  313000}
\address{$^{22}$Indian Institute of Science Education and Research (IISER), Berhampur 760010 , India}
\address{$^{23}$Indian Institute of Science Education and Research (IISER) Tirupati, Tirupati 517507, India}
\address{$^{24}$Indian Institute Technology, Patna, Bihar 801106, India}
\address{$^{25}$Indiana University, Bloomington, Indiana 47408}
\address{$^{26}$Institute of Modern Physics, Chinese Academy of Sciences, Lanzhou, Gansu 730000 }
\address{$^{27}$University of Jammu, Jammu 180001, India}
\address{$^{28}$Joint Institute for Nuclear Research, Dubna 141 980, Russia}
\address{$^{29}$Kent State University, Kent, Ohio 44242}
\address{$^{30}$University of Kentucky, Lexington, Kentucky 40506-0055}
\address{$^{31}$Lawrence Berkeley National Laboratory, Berkeley, California 94720}
\address{$^{32}$Lehigh University, Bethlehem, Pennsylvania 18015}
\address{$^{33}$Max-Planck-Institut f\"ur Physik, Munich 80805, Germany}
\address{$^{34}$Michigan State University, East Lansing, Michigan 48824}
\address{$^{35}$National Research Nuclear University MEPhI, Moscow 115409, Russia}
\address{$^{36}$National Institute of Science Education and Research, HBNI, Jatni 752050, India}
\address{$^{37}$National Cheng Kung University, Tainan 70101 }
\address{$^{38}$Nuclear Physics Institute of the CAS, Rez 250 68, Czech Republic}
\address{$^{39}$Ohio State University, Columbus, Ohio 43210}
\address{$^{40}$Institute of Nuclear Physics PAN, Cracow 31-342, Poland}
\address{$^{41}$Panjab University, Chandigarh 160014, India}
\address{$^{42}$Pennsylvania State University, University Park, Pennsylvania 16802}
\address{$^{43}$NRC "Kurchatov Institute", Institute of High Energy Physics, Protvino 142281, Russia}
\address{$^{44}$Purdue University, West Lafayette, Indiana 47907}
\address{$^{45}$Rice University, Houston, Texas 77251}
\address{$^{46}$Rutgers University, Piscataway, New Jersey 08854}
\address{$^{47}$Universidade de S\~ao Paulo, S\~ao Paulo, Brazil 05314-970}
\address{$^{48}$University of Science and Technology of China, Hefei, Anhui 230026}
\address{$^{49}$Shandong University, Qingdao, Shandong 266237}
\address{$^{50}$Shanghai Institute of Applied Physics, Chinese Academy of Sciences, Shanghai 201800}
\address{$^{51}$Southern Connecticut State University, New Haven, Connecticut 06515}
\address{$^{52}$State University of New York, Stony Brook, New York 11794}
\address{$^{53}$Instituto de Alta Investigaci\'on, Universidad de Tarapac\'a, Chile}
\address{$^{54}$Temple University, Philadelphia, Pennsylvania 19122}
\address{$^{55}$Texas A\&M University, College Station, Texas 77843}
\address{$^{56}$University of Texas, Austin, Texas 78712}
\address{$^{57}$Tsinghua University, Beijing 100084}
\address{$^{58}$University of Tsukuba, Tsukuba, Ibaraki 305-8571, Japan}
\address{$^{59}$United States Naval Academy, Annapolis, Maryland 21402}
\address{$^{60}$Valparaiso University, Valparaiso, Indiana 46383}
\address{$^{61}$Variable Energy Cyclotron Centre, Kolkata 700064, India}
\address{$^{62}$Warsaw University of Technology, Warsaw 00-661, Poland}
\address{$^{63}$Wayne State University, Detroit, Michigan 48201}
\address{$^{64}$Yale University, New Haven, Connecticut 06520}

\begin{abstract}
We present a measurement of the first-order azimuthal anisotropy, $v_1$, of deuterons from Au+Au collisions at
$\ems$ = 7.7, 11.5, 14.5, 19.6, 27, and 39 GeV recorded with the STAR experiment at the Relativistic Heavy Ion Collider (RHIC).
The energy dependence of the $\vone$ slope, $\dvdy|_{y=0}$, for deuterons, where $y$ is the rapidity, is extracted for semi-central
	collisions (10-40\% centrality) and compared with that of protons. 
While the $\vone$ slopes of protons are generally negative for $\ems >$ 10 GeV,
those for deuterons are consistent with zero, a strong enhancement of the $\vone$ slope of deuterons is seen at the lowest
	collision energy (the largest baryon density) at $\ems =$ 7.7 GeV.
In addition, we report the transverse momentum dependence of $v_1$ for protons and deuterons.
The experimental results are compared with transport and coalescence models.

\end{abstract}

\maketitle

\hyphenpenalty=700
% \tolerance=100
%%%%%%%%%%%%%%%%%%%%%%%%%%%%%%%%%%%%%%%%%%%%%%%%%%%%%%%%%%%%%%%%%%%%%%%%%%%%%%%%%%%%%%%%%%%%%%%%%%%%%%%%%%
\section{Introduction}
 One of the main goals of high-energy heavy-ion collision experiments is to explore the state and evolution of nuclear matter
 under extreme conditions.
 These experiments measure the multiplicities of many different particle species and the correlations between these particles.
 The correlations between the azimuthal angles of these particles are particularly informative.
 The directed flow, $v_1$, and the elliptic flow, $v_2$, are the first and second harmonic coefficients of the Fourier
 expansion of the particle azimuthal distributions in momentum space relative to the reaction-plane~\cite{event_plane}. The reaction-plane is defined by the beam direction and the impact parameter.
 The directed flow has two components: a rapidity-even function, $v_1^{\rm even}$, and a rapidity-odd function, $v_1^{\rm odd}$.
 The values of $v_1^{\rm even}$ represent the contribution from event-by-event initial nuclei geometry fluctuations~\cite{Teaney_2011,Luzum_2011}.
 This work will focus on the rapidity-odd component.
 The values of $v_1$ as a function of rapidity, $y$, are sensitive to the amount of expansion the collision system goes through
 during the early collision stages~\cite{Bozek_2010}.

 The RHIC has completed the first phase of the Beam Energy Scan (BES) program~\cite{bes-I}.
 The directed flow $v_1(y)$ as a function of rapidity, $y$, for different mesons and baryons has been measured in Au+Au collisions over the range of beam energies of
 $\ems = $ 7.7 to 200 GeV~\cite{Adamczyk_2014,Adamczyk_2018}.
 The slopes $\dvdy|_{y=0}$ at mid-rapidity for net-protons and net-$\Lambda$ hyperons as a function of collision energy show a minimum around $\ems$ = 10-20 GeV.
 According to a hydrodynamic model~\cite{Stocker_2005}, a minimum $\dvdy|_{y=0}$ of net-baryons as a function of collision energy is a signature of a first-order
 phase transition between hadronic matter and the quark gluon plasma.
 However, no existing hydrodynamic model can quantitatively reproduce the measured
 magnitudes of the meson and baryon directed flow~\cite{Adamczyk_2014,Adamczyk_2018}.

 Besides the charged hadrons, a large number of light nuclei are produced in heavy-ion collisions.
 Their production is sensitive to the properties of cluster formation and fireball evolution ~\cite{Kupasta_1980,Mekjian_1978,Csernai_1985,Mattiello_1997,Sun_2017}.
 There are two commonly-used and very different phenomenological pictures for the mechanisms governing the production of light nuclei.
 The thermal model describes deuteron production as occurring throughout the whole time evolution of the fireball up to chemical freeze-out
 via elementary nucleon-nucleon and/or parton-parton interactions~\cite{Mekjian_1978,Munzinger2_1995,Chatterjee_2014}.
 Such models are able to reproduce the observed deuteron multiplicities~\cite{Andronic_2011,Cleymans_2011}.
 It is, however, difficult to understand how deuterons formed in the intermediate stages of the collisions can survive the subsequent evolution, as their binding energy (2 MeV) is so small compared with the fireball temperature ($\approx$150 MeV~\cite{Adamczyk_2017}).
 Another model describes deuteron production as occurring much later in the collision,
 near kinetic freeze-out, when the temperatures are much lower~\cite{Butler_1963,Gutbrod_1976,Sato_1981,Zhang_2010,Steinherimer_2012}.
 This is the coalescence model, in which two nucleons that are near each other in space and traveling with similar velocities, can form a deuteron.
 Thus, the momentum distribution of these formed deuterons is strongly related to that of protons.
 The comparison of light nucleus directed flow with that of protons can provide additional information
 to understand the mechanisms involved in light nucleus production in high-energy heavy-ion collisions.

 Both the EOS and FOPI collaborations observed energy dependence of the directed flow for protons and deuterons from Au+Au collisions
 for lab kinetic energies of $0.1A$ GeV to $1.5A$ GeV~\cite{Partlan_1995, Wang_1995,Reisdorf_2012}.
 These observations suggest that the directed flow of deuterons has a more pronounced energy dependence than that of protons. Thus, the light nucleus directed flow may provide a more sensitive measure of the collective motion than the lighter hadrons.

 In this paper, we present the measurement of the directed flow for deuterons in Au+Au collisions at $\ems$ = 7.7, 11.5, 14.5, 19.6, 27, and 39 GeV from the STAR experiment.
 The results are discussed and compared with AMPT (A Multi-Phase Transport) calculation~\cite{Lin_2005} and a simple coalescence model~\cite{Molnar_2003}.

%%%%%%%%%%%%%%%%%%%%%%%%%%%%%%%%%%%%%%%%%%%%%%%%%%%%%%%%%%%%%%%%%%%%%%%%%%%%%%%%%%%%%%%%%%%%%%%%%%%%%%%%%%
\section{Experiment and Data Analysis}
 The data used here are for Au+Au collisions at beam energies of $\ems$ = 7.7, 11.5, 14.5,
 19.6, 27, and 39 GeV collected by the STAR experiment~\cite{star} at the RHIC facility.
 A minimum bias trigger was used.
 The 7.7, 11.5, and 39 GeV data were recorded in 2010.
 The 19.6 and 27 GeV data were recorded in 2011, and the 14.5 GeV data were recorded in 2014.
 The STAR experiment consists of a solenoidal magnet and different detectors
 for tracking, triggering, and particle identification (PID).
 The Time Projection Chamber (TPC)~\cite{tpc} is a
 charged-particle tracking device which covers the full azimuth and a pseudo-rapidity range $|\eta| < 1$.
 Charged particle trajectories are reconstructed with the TPC, and the momentum components are obtained from the curvature of the helical path in the 0.5 Tesla magnetic field.
 The two momentum components in the plane transverse to the beam-line define the azimuthal angle of each track.
 The main detectors used for PID are the TPC and the Time-of-Flight system (TOF)~\cite{tof}.
 The details of other STAR detectors are described elsewhere~\cite{star}.

%%%%%%%%%%%%%%%%%%%%%%%%%%%%%%%%%%%%%%%%%%%%%%%%%%%%%%%%%%%%%%%%%%%%%%%%%%%%%%%%%%%%%%%%%%%%%%%%%%%%%%%%%%
\subsection{Event and Track Selection}
 For each event, the location of the primary vertex can be reconstructed in three dimensions by extrapolating the TPC track segments to the beam-axis.
 The primary vertex is required to be within certain distances of the center of STAR
 in the directions along the beam axis, $v_z$, and transverse to it, $v_r$, as listed in Table~\ref{table1}. 
\begin{table}[h]
\centering
\caption{The event selection quality cuts $v_z$ and $v_r$ (see text), the
number of events, and the baryon chemical potential, $\mu_B$~\cite{Cleymans_2006}, at each of the different collision energies studied here.
The center of transverse radial position is
located at ($v_x$, $v_y$) = (0, -0.89 cm) for 14.5 GeV.}
\label{table1}
\begin{tabular}{|p{1.8cm}<{\centering}|m{1.2cm}<{\centering}|m{1.2cm}<{\centering}|m{1.8cm}<{\centering}|p{1.5cm}<{\centering}|}
\hline
$\sqrt{s_{NN}}$ (GeV)  & $|v_{z}|$ (cm)& $v_{r}$ (cm) & Events ($\times 10^{6}$) &$\mu_B $(MeV) \\
\hline
7.7 & 70 & 2  & 4 & 420\\
11.5 & 50 & 2  & 12 & 315 \\
14.5 & 50 & 1 & 11 & 260\\
19.6 & 50 & 2  & 36 & 205\\
27 & 50 & 2  & 70 & 155\\
39 & 40 & 2  & 130 & 115 \\
\hline
\end{tabular}
\end{table}

 The reconstructed tracks used in this analysis were required to pass basic quality cuts, including having at least 15 TPC space points assigned to them.
 Each track is also required to extrapolate to within 1 cm of the primary vertex location (distance of closest approach DCA), and has
 assigned to it at least half of the possible number of TPC space points (maximum 45) for its trajectory.

 The centrality of each event is determined by comparing the charged particle multiplicity measured in the event to a Monte-Carlo Glauber reference~\cite{Adamczyk_2012}.
 The results presented in this paper use the 10-40\% intermediate
 centrality region where the $v_1(y)$ measurements are the most significant.
 The first-order event plane resolution, and $v_1$ itself, in more central collisions are relatively smaller,
 while the deuteron yields are
 also relatively smaller in more peripheral collisions.

%%%%%%%%%%%%%%%%%%%%%%%%%%%%%%%%%%%%%%%%%%%%%%%%%%%%%%%%%%%%%%%%%%%%%%%%%%%%%%%%%%%%%%%%%%%%%%%%%%%%%%%%%%
\subsection{Particle Identification}
\begin{figure}[htbp]
\centering
\includegraphics[width=8.5cm]{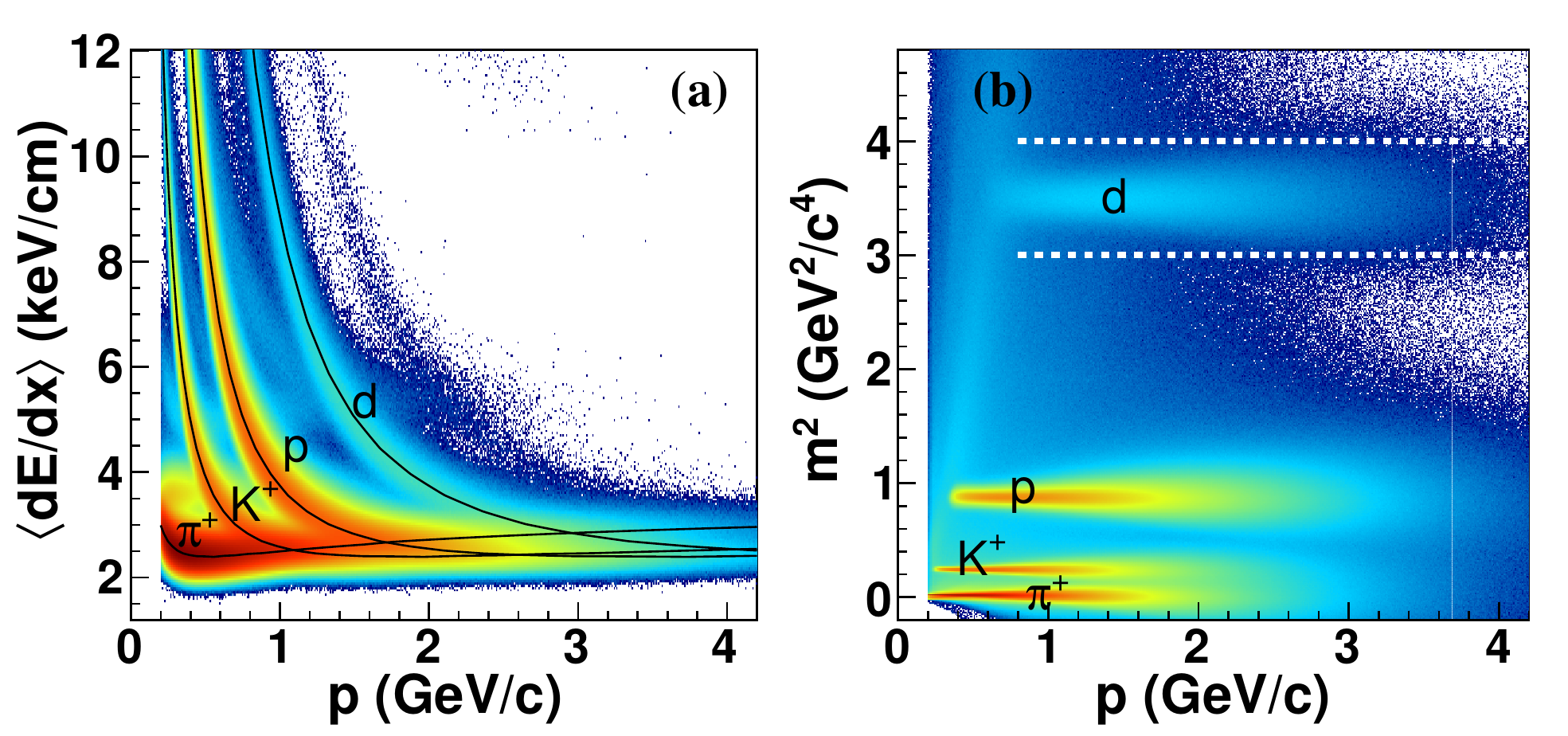}
\caption{(a) The $\langle{dE/dx}\rangle$ of charged tracks versus momentum in Au+Au collisions at $\ems = $ 19.6 GeV.
The curves are Bichsel predictions for the corresponding particle species.
(b) Particle $m^2$ versus momentum at $\ems = $ 19.6 GeV.
The bands, from bottom to top, correspond to $\pi^+$, $K^+$, protons, and deuterons, respectively.
}\label{fig_pid}
\end{figure}

 We use a combination of the TPC and the TOF for the identification of charged particles.
 Figure~\ref{fig_pid}(a) shows the average $dE/dx$ distribution of measured charged tracks versus momentum at $\ems = $ 19.6 GeV.
 The curves denote the Bichsel expectation values, $\langle{dE/dx}_{B}\rangle$, for each species~\cite{Bichsel}.

 For each track, the particle speed divided by the speed
 of light, $\beta=v/c$, can be measured by the combination of the TPC and TOF systems.
 The TOF thus provides a measurement of the track mass-squared, $m^2$, according to
 \begin{equation}\label{eq_m2}
    m^2 = p^2\biggl(\frac{1}{{\beta}^2}-1 \biggr),
 \end{equation}
 where $p$ is the track momentum measured in the TPC.
 Figure~\ref{fig_pid}(b) shows the $m^2$ distribution as a function of momentum at $\ems = $ 19.6 GeV.
 For the deuteron selection, the mass-squared values are required to be in the range 3.0 GeV$^2$/$c^4$ $ < m^2 < $ 4.0 GeV$^2$/$c^4$.

\begin{figure}[htbp]
\centering
\includegraphics[width=9cm]{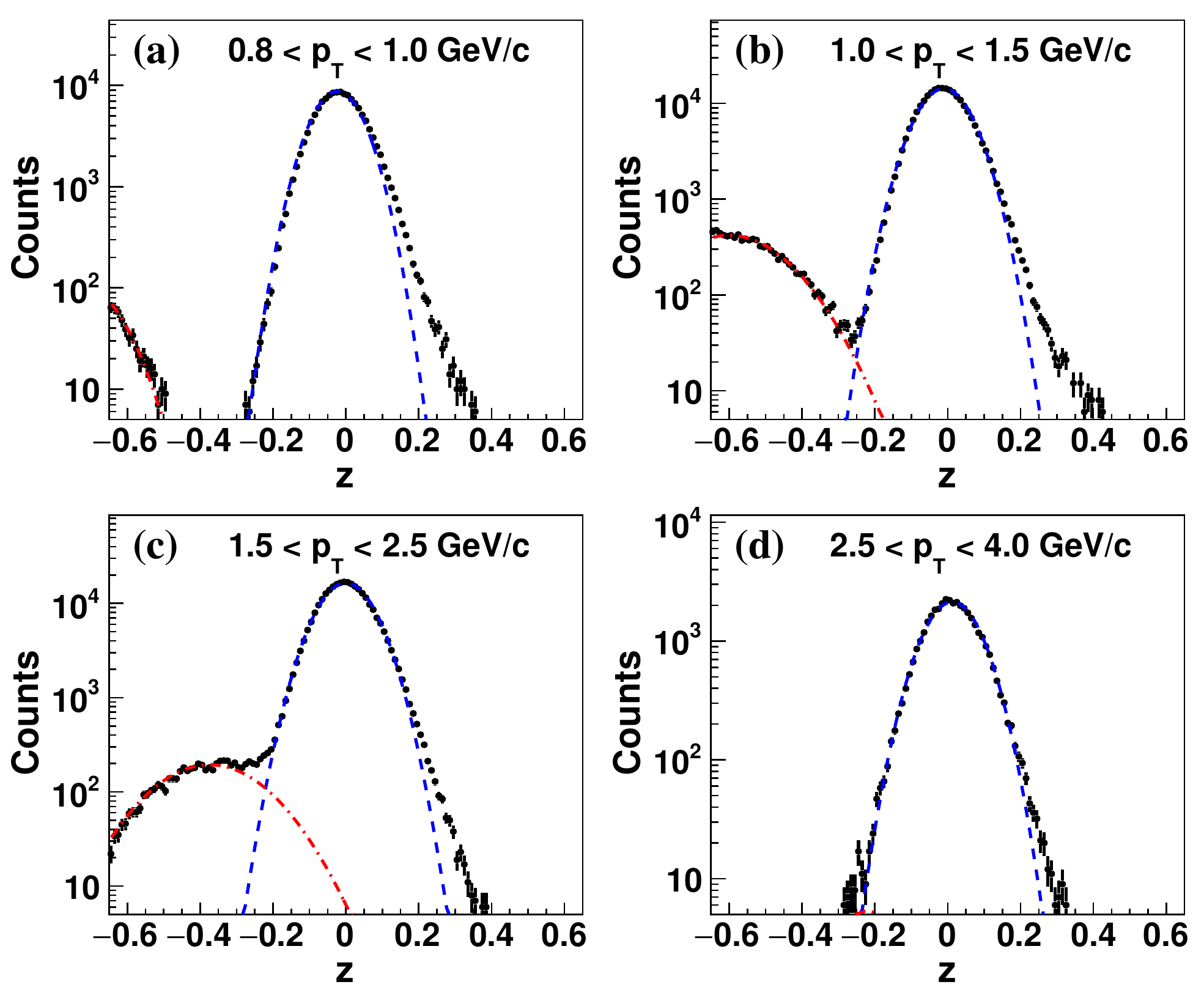}
\caption{The $z$ distribution for deuteron in various $p_{\rm T}$ ranges in Au+Au
collisions at $\ems = $ 19.6 GeV. The dashed line is a Gaussian fit representing the distribution for deuterons.
The dot-dashed curve is a Gaussian fit denoting contributions from $\pi^{+}$, $K^{+}$, and protons.
}\label{fig_z}
\end{figure}

 The selection of deuteron tracks using the TPC $\langle{dE/dx}\rangle$ proceeds via
 the variable $z$, defined as~\cite{Adam_2019},
\begin{equation}\label{eq_z}
 z=\ln\biggl(\frac{\langle{dE/dx}\rangle}{\langle{dE/dx}_{B}\rangle}\biggr).
\end{equation}
 When using the Bichsel prediction, $\langle{dE/dx}_{B}\rangle$, for deuterons in Eq.~\ref{eq_z} ({\it cf.} Fig.~\ref{fig_pid}),
 the deuterons are those tracks with values of $z$ near zero.
 Figure~\ref{fig_z} shows the $z$ distributions in different $p_{\rm T}$ ranges at $\ems = $ 19.6 GeV.
 In this analysis, the deuteron selection involves the requirement that $|z| <$ 0.2.

 In Ref.~\cite{Adamczyk_2014}, the $\vone$ of protons was measured over the range of 0.4 GeV/$c$ $< p_{\rm T} <$ 2.0 GeV/$c$.
 For the deuterons in this analysis, the transverse momentum range is restricted to the same range in terms of $p_{\rm T}$/A, or 0.8 GeV/$c$ $< p_{\rm T} <$ 4.0 GeV/$c$.
 The default rapidity window for extracting the $\vone$ slope is $|y| < 0.6$.

%%%%%%%%%%%%%%%%%%%%%%%%%%%%%%%%%%%%%%%%%%%%%%%%%%%%%%%%%%%%%%%%%%%%%%%%%%%%%%%%%%%%%%%%%%%%%%%%%%%%%%%%%%
\subsection{Event Plane}
 The reaction-plane angle, $\Psi_R$, is the azimuth of the plane spanned by the beam direction and the impact parameter vector.
 The $v_1$ of the produced particles with respect to $\Psi_R$ can be measured as $v_1 = \langle\cos(\phi-\Psi_R)\rangle$,
 where $\phi$ is the azimuthal angle of the produced particle and the angle brackets imply averaging over all the particles in all events.
 As the reaction-plane angle, $\Psi_R$, cannot be measured directly, we will use the event-plane angle~\cite{event_plane}
 to estimate the reaction-plane angle $\Psi_R$.
 The event-plane was estimated using the $v_1$ information of the final-state particles, and hence is
 called the first-order event-plane ($\Psi_1$).
 The self-correlations were eliminated with the large acceptance gap between the TPC,
 where the deuteron directed flow was measured, and the detectors measuring the final-state particles used to calculate $\Psi_1$.

\begin{figure}[htbp]
\centering
\includegraphics[width=8.5cm]{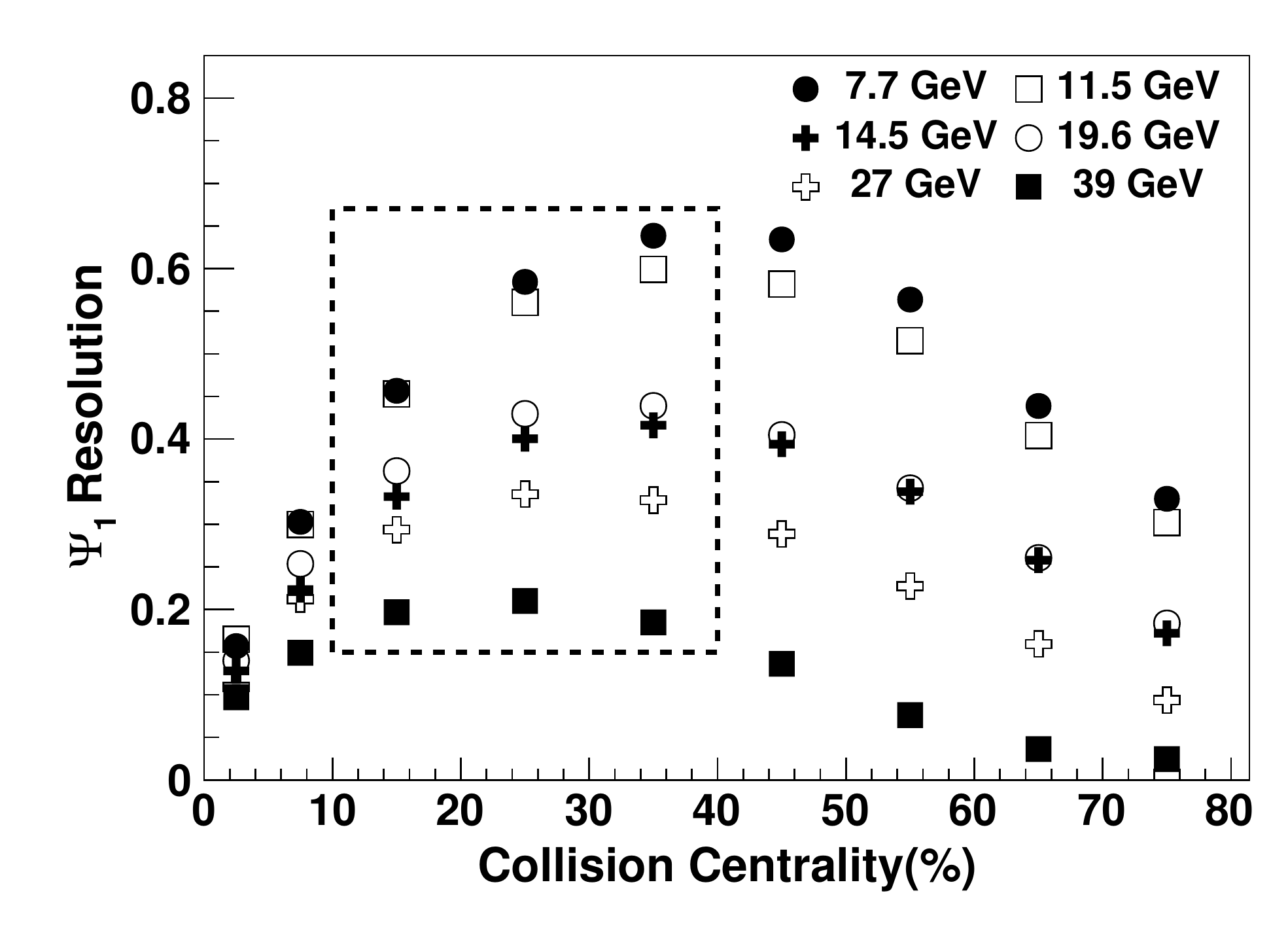}
\caption{The values of the first-order event plane ($\Psi_{1}$) resolution $R_1$ as a function of the centrality of Au+Au collisions at
$\ems$ = 7.7, 11.5, 14.5, 19.6, 27, and 39 GeV.
The $\Psi_{1}$ was reconstructed with the BBCs and its resolution is determined by the correlation of the sub-event-plane angles determined separately by the east and west BBCs.
Data presented later (10-40\% centrality) are indicated by the dashed-line box.
}\label{fig_RP}
\end{figure}

 Two beam-beam counters (BBCs)~\cite{bbc} were used to reconstruct the values of $\Psi_1$.
 The distribution of reconstructed $\Psi_1$ values is not uniform due to imperfections in the BBCs.
 Therefore, a shifting method~\cite{event_plane} was applied
 to flatten the distributions.
 The finite multiplicity of particles in each event limits the precision of estimating the true reaction-plane via the reconstructed
 $\Psi_1$, so the values of $v_1$ have been corrected for the event plane resolution : $v_1 = \langle\cos(\phi-\Psi_1)\rangle/R_1$.
 The resolution correction factor, $R_1$, is determined by the sub-event plane correlation method~\cite{event_plane},
 where the sub-event planes are reconstructed separately in the east and west BBCs.
 Figure~\ref{fig_RP} shows the $R_1$ values as a function of the collision centrality at each collision energy.
 The resolution peaks in mid-central collisions.
 The resolution improves as the collision energy decreases due to the stronger directed flow at the rapidity ranges covered by the BBC detectors.

%%%%%%%%%%%%%%%%%%%%%%%%%%%%%%%%%%%%%%%%%%%%%%%%%%%%%%%%%%%%%%%%%%%%%%%%%%%%%%%%%%%%%%%%%%%%%%%%%%%%%%%%%%
\subsection{Systematic Uncertainties}
 The systematic uncertainties on the directed flow are estimated by varying the criteria used to select tracks and identify particles.
 The absolute difference between the results using the default and the varied criteria is quoted as the systematic uncertainty.
 The systematic uncertainty related to the track selection procedure is estimated by varying the DCA (from 1 to 0.5 and 2 cm) and the number of TPC space points (from 15 to 20). The value is found to be about $6\%$ and independent of the particle species.
 Additional systematic uncertainties arising from the particle
 misidentification are estimated by varying the PID cuts on $z$ and $m^2$.
 The typical magnitudes of these uncertainties are $4\%$ for protons and $15\%$ for deuterons.
 The systematic uncertainty corresponding to the chosen range of the $dv_1/dy$
 fit is estimated by taking the difference between the best fitted slope, and
 the value of the slope within $|y| < 0.5$. This uncertainty is about $6\%$ at $\ems=$ 7.7 GeV.
 It is the choice of the $dv_1/dy$ fit range that makes the largest contribution to the total systematic uncertainties above 7.7 GeV.
 Non-flow contributions to the systematic uncertainty are reduced due to the large pseudo-rapidity gap between the TPC and BBC detectors.
 The event-plane resolution is estimated via the correlation of the
 event-planes calculated for two sub-events, 
 which can be affected by momentum conservation~\cite{Borghini_2002}.
 The possible systematic uncertainty from the first-order event-plane
 resolution estimation is discussed in Ref.~\cite{Adamczyk_2014}. The uncertainty is less than $2\%$.
 All the sources are added in quadrature as the final total systematic uncertainties, which are of a similar magnitude as the statistical uncertainties.

%%%%%%%%%%%%%%%%%%%%%%%%%%%%%%%%%%%%%%%%%%%%%%%%%%%%%%%%%%%%%%%%%%%%%%%%%%%%%%%%%%%%%%%%%%%%%%%%%%%%%%%%%%
\section{Results and Discussion}

\begin{figure}[htbp]
\centering
\includegraphics[width=9cm]{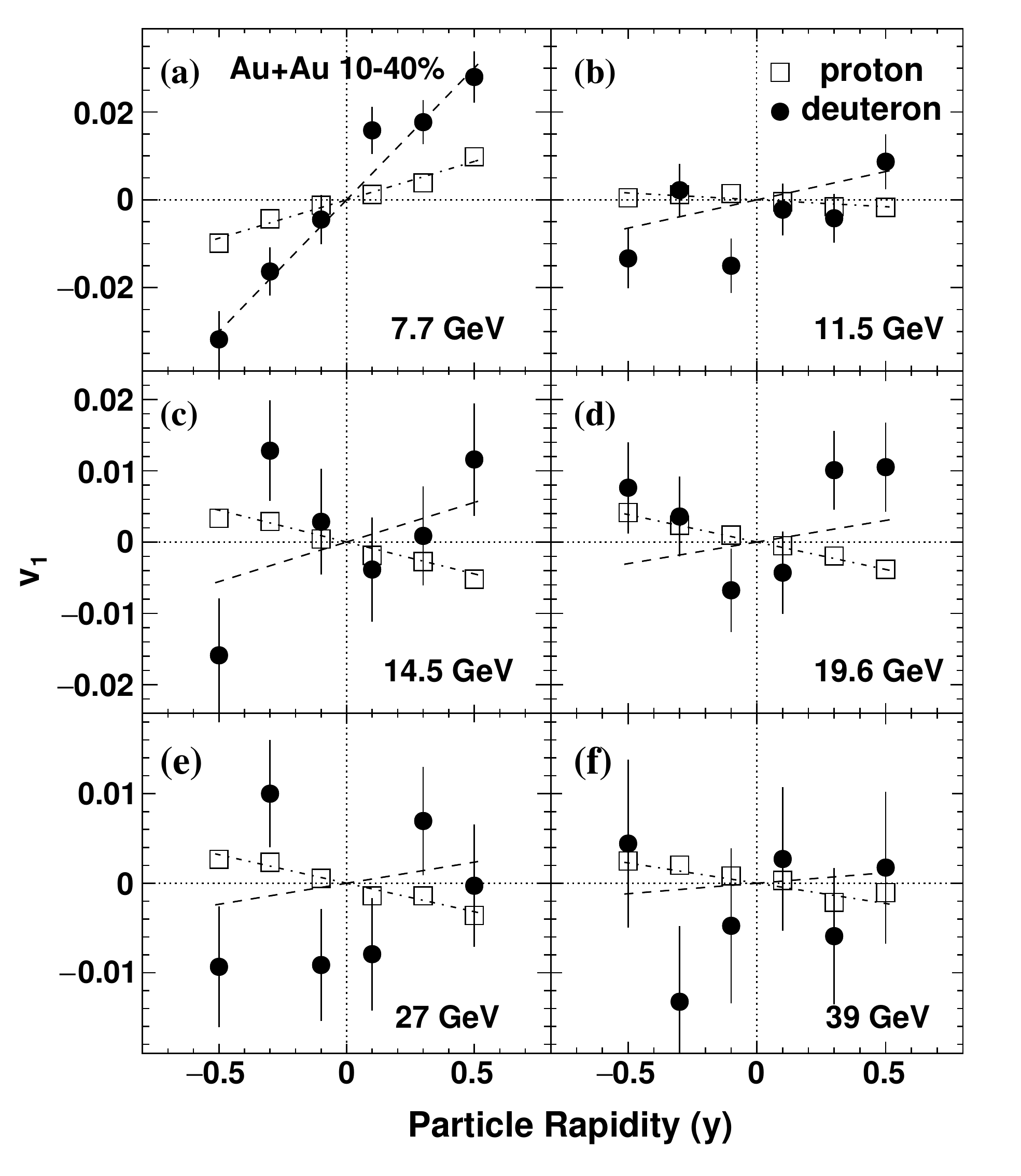}
\caption{Rapidity dependence of $v_1$ for protons~\cite{Adamczyk_2018} (open squares) and deuterons (solid circles) in 10-40\% Au+Au collisions at $\ems$ = 7.7, 11.5, 14.5, 19.6, 27, and 39 GeV.
The dot-dashed and dashed lines are fits to proton and deuteron $v_1$, respectively, at midrapidity ($|y| < 0.6$) with a linear function to extract the slopes.
The plotted uncertainties are statistical only.
}\label{fig_v1y}
\end{figure}

 Figure~\ref{fig_v1y} shows the rapidity dependence of the directed flow, $v_1$, of protons and deuterons at each of the studied collision energies.
 The $\vone$ of deuterons is antisymmetric about $y=0$. As with the protons, the $\vone$ of deuterons
  increases monotonically with increasing rapidity at $\ems=$ 7.7 GeV.
 We observe a stronger $v_1$ dependence on rapidity for deuterons than for protons.
 The limited event statistics and relatively lower deuteron
 production rate at higher energies makes such comparisons less certain.

\begin{figure}[ht]
\centering
\includegraphics[width=8.5cm]{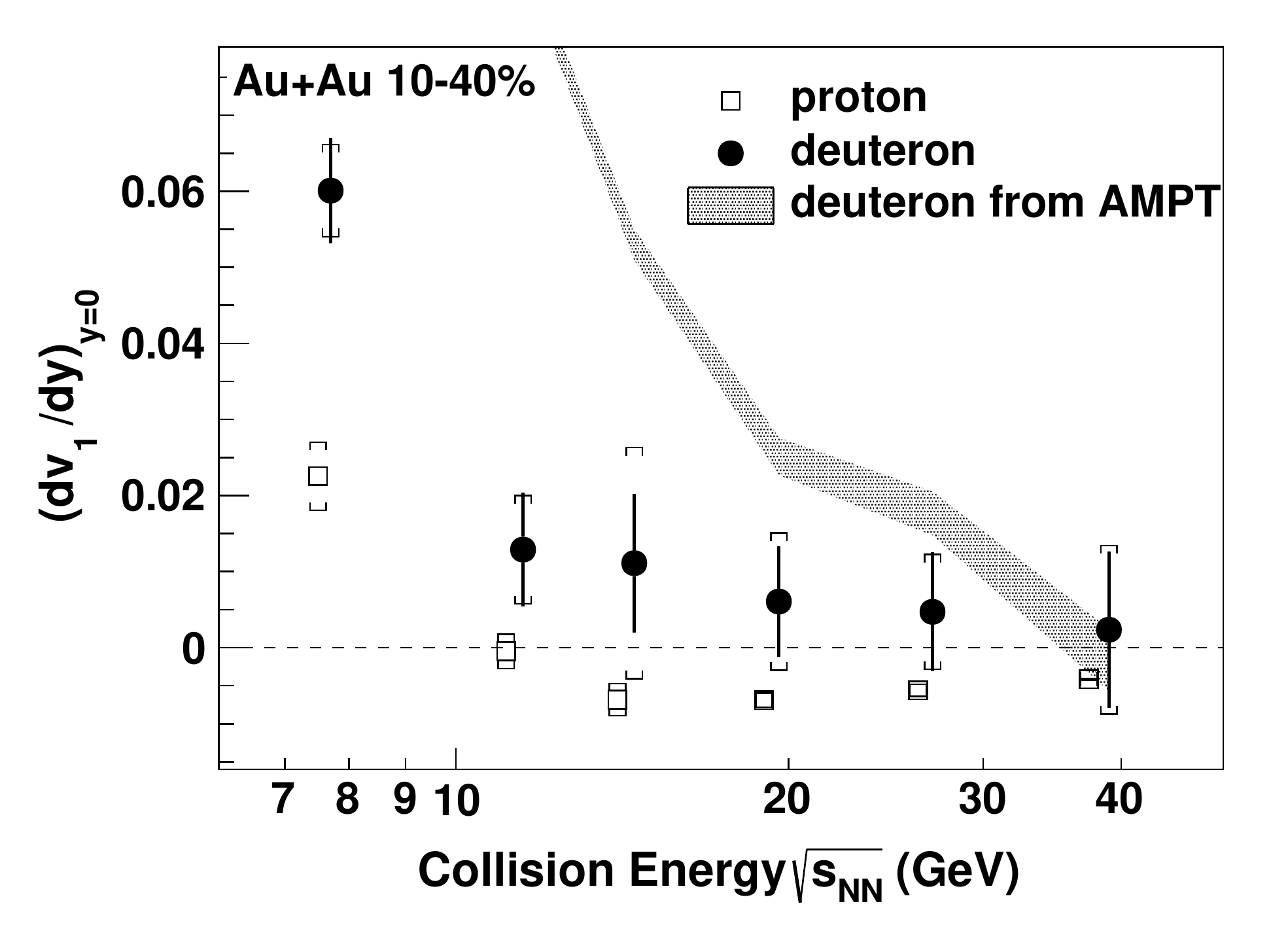}
\caption{Directed flow slope at mid-rapidity, $\dvdy|_{y=0}$, as a function of beam energy in 10-40\% Au+Au collisions.
Solid circles represent deuterons. Open squares are the published results in~\cite{Adamczyk_2018} for protons.
The band denotes the results for deuterons from AMPT transport model.
Statistical uncertainties (bars) and systematic uncertainties (horizontal brackets) are shown separately.
For visibility, the data points are staggered horizontally.
}\label{fig_dv1dy}
\end{figure}

 The $\vone$ slope at mid-rapidity ($y<|0.6|$) is obtained by fitting the data with a straight line.
 For $\ems >$ 7.7 GeV, the sign of the $\vone$ slope is mainly influenced
 by the two data points at the extreme rapidity bins.
 Figure~\ref{fig_dv1dy} presents the resulting values of the $\vone$ slope versus the collision energy for 10-40\% central collisions.
 A significantly larger deuteron $\vone$ slope with respect to protons is observed at $\ems =$ 7.7 GeV.
 The deuteron $\vone$ slope is observed to be consistent with zero at all energies above 7.7 GeV,
 but with large uncertainties.

 The results from the data were compared to those
 from the AMPT model~\cite{Lin_2005}.
 This is a hybrid model which has been used to describe the charged particle multiplicity,
 transverse momentum, and the elliptic flow of identified particles in relativistic heavy-ion collisions.
 In this model, scattering among hadrons is described by ART (A Relativistic Transport) model~\cite{Li_1995}.
 The deuterons are produced and dissolved within ART via nuclear reactions.
 The centrality of the simulated events is determined by integrating the charged particle multiplicity distribution, as was done for the experimental data.
 The comparison between data and the AMPT model result can be seen in Fig.~\ref{fig_dv1dy}.
 A decreasing trend for increasing collision energies is seen in the AMPT simulation,
 while the model significantly overpredicts the observed magnitude of the deuteron directed flow slope.

 A commonly-applied picture for light nucleus production in heavy-ion collisions involves the coalescence of nucleons which are close to each other in space and have similar velocities.
 Then, the spectral distribution of a light nucleus, $d^{3}N_{A}/d^{3}p_{A}$, depends on the distributions of protons, $d^{3}N_{p}/d^{3}p_{p}$, and neutrons, $d^{3}N_{n}/d^{3}p_{n}$,~\cite{Sato_1981},
 \begin{equation}\label{eq_coalescence}
   E_{A}\frac{d^{3}N_{A}}{d^{3}p_{A}}\propto\Bigl(E_{p}\frac{d^{3}N_{p}}{d^{3}p_{p}}\Bigr)^{Z}
   \Bigl(E_{n}\frac{d^{3}N_{n}}{d^{3}p_{n}}\Bigr)^{A-Z},
 \end{equation}
 where A and Z are nucleus mass number and charge number, respectively.
 In this production mechanism, the expected value of the light nucleus directed flow can be expressed as a function of the directed flow of its constituent nucleons.
 Assuming the protons and neutrons flow similarly, the deuteron $v_1$ is given by~\cite{Molnar_2003}:
 \begin{equation}\label{eq_dv1}
   v_{1,d}(y, p_{\rm T})=\frac{2v_{1,p}(y, \frac{p_{\rm T}}{2})}{1+2v^2_{1,p}(y, \frac{p_{\rm T}}{2})},
 \end{equation}
 where each constituent nucleon has half the $p_{\rm T}$ and the same rapidity as the deuteron.
 Then one can calculate the expected $v_1$ for the deuterons from the measured $v_1$ for protons~\cite{Adamczyk_2018},
 assuming as usual the (unmeasured) neutron flow is the same as that of the (measured) protons.
 As the proton $v_1 \ll 1$, Eq.~\ref{eq_dv1} can be simplified as
\begin{equation}\label{eq_dv1_1}
   v_{1,d}(y, p_{\rm T})\approx2v_{1,p}(y, \frac{p_{\rm T}}{2}).
\end{equation}
 This indicates that, in the coalescence mechanism, the $v_1$ of protons and deuterons will follow an atomic mass-number scaling.
 In fact, Eq.~\ref{eq_dv1} and Eq.~\ref{eq_dv1_1} can be applied to any anisotropy coefficient.
 For the elliptic flow of light nuclei, the STAR collaboration has observed such a mass-number scaling
 in $\ems=$ 7.7-200 GeV Au+Au collisions~\cite{Adamczyk_2016}.
 The expectation would thus be that the $\vone$ slope for
 deuterons would have the same sign as that observed for the protons and have a larger magnitude.
 In Fig.~\ref{fig_dv1dy}, within the statistical and systematic uncertainties,
 the deuteron $\vone$ slope at mid-rapidity is consistent with this expectation at $\ems=$ 7.7 GeV.
 For $\ems>$7.7 GeV, the deuteron $\vone$
 slopes have a different sign than the corresponding proton $\vone$ slopes
 with large uncertainties.

\begin{figure}
\centering
\includegraphics[width=9cm]{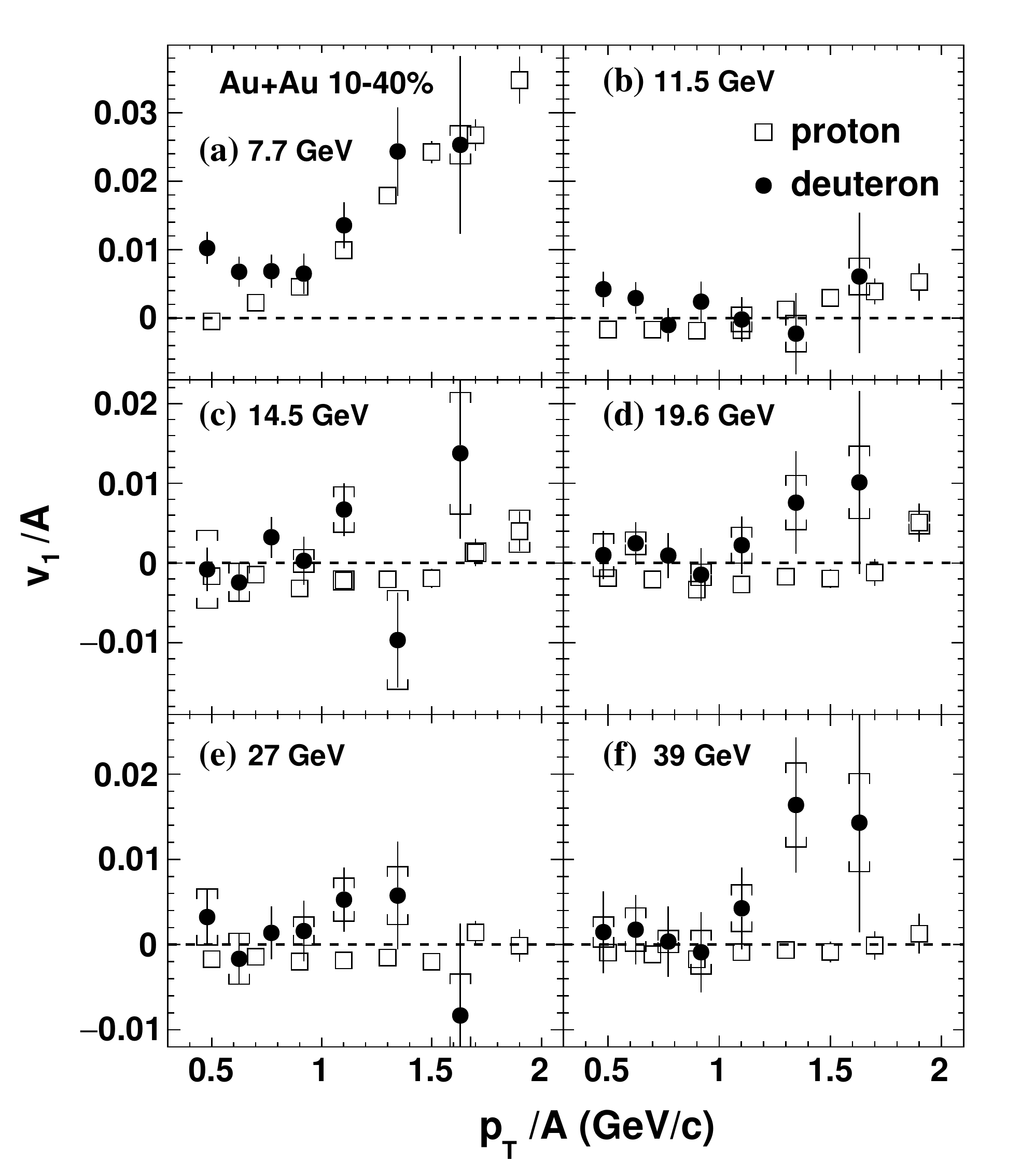}
\caption{The $p_{\rm T}$ dependence of $v_1/A$ in $|y|<0.6$ for protons
(open squares) and deuterons (solid circles) in 10-40\% Au+Au collisions at $\ems$ = 7.7, 11.5, 14.5, 19.6, 27, and 39 GeV.
Statistical uncertainties (bars) and systematic uncertainties (horizontal lines) are shown separately.
}\label{fig_v1pt}
\end{figure}

 To further test the coalescence model, we studied the $p_{\rm T}$ dependence of the directed flow, $v_1$, at all measured energies, which is shown in Fig.~\ref{fig_v1pt}.
 At $\ems=$ 7.7 GeV and 11.5 GeV, the values of $v_1(p_{\rm T})$ indicate a mass-number scaling for $p_{\rm T}/A > 1$ GeV/$c$ within $|y|<0.6$, while the value of the deuteron $v_1/A$ shows an enhancement towards lower $p_{\rm T}/A$
 at $\ems=$ 7.7 GeV. This enhancement is not caused by the knock-out deuteron background with its negligible production at $\ems=$ 7.7 GeV~\cite{Adam_2019}.

 At forward rapidities, the E877 collaboration observed such an enhancement of the $v_1(p_{\rm T})$ of deuterons, tritons, ${\rm ^3He}$, and ${\rm ^4He} $
 at $p_{\rm T} < 0.5$ GeV/$c$ in Au+Au collisions at a beam energy of 10.8$A$ GeV~\cite{Barrette_1999}.
 The cause of the low $p_{\rm T}$ enhancement of the deuteron $v_1$ in the $\ems = $ 7.7 GeV Au+Au collisions is unclear.

%%%%%%%%%%%%%%%%%%%%%%%%%%%%%%%%%%%%%%%%%%%%%%%%%%%%%%%%%%%%%%%%%%%%%%%%%%%%%%%%%%%%%%%%%%%%%%%%%%%%%%%%%%
\section{Summary}
In summary, we present the mid-rapidity directed flow $\vone$ of deuterons in Au+Au collisions at $\ems =$ 7.7-39 GeV.
At 10-40\% centrality, the $\vone$ slope, $\dvdy|_{y=0}$,
shows a strong increase at the lowest collision energy of $\ems =$ 7.7 GeV, and is consistent with zero for energies above 7.7 GeV.
The AMPT transport model significantly overestimates the values of deuteron $\vone$ slopes at most measured collision energies.
The coalescence model for deuteron production predicts an atomic-mass-number scaling of the proton and deuteron $v_1$.
At $\ems =$ 7.7 GeV and 11.5 GeV, this is approximately valid for the $v_1({p_{\rm T}}$) data at higher
$p_{\rm T}$ within $|y| <$ 0.6, while the $v_1({p_{\rm T}}$) show enhancements towards very low $p_{\rm T}$ at $\ems =$ 7.7 GeV. There is at present no explanation for this enhancement.
Stronger conclusions will be possible with the event statistics achieved with the Beam Energy Scan II program.

\section*{Acknowledgement}
We thank the RHIC Operations Group and RCF at BNL, the NERSC Center at LBNL, and the Open Science Grid consortium for providing resources and support.  This work was supported in part by the Office of Nuclear Physics within the U.S. DOE Office of Science, the U.S. National Science Foundation, the Ministry of Education and Science of the Russian Federation, National Natural Science Foundation of China, Chinese Academy of Science, the Ministry of Science and Technology of China and the Chinese Ministry of Education, the Higher Education Sprout Project by Ministry of Education at NCKU, the National Research Foundation of Korea, Czech Science Foundation and Ministry of Education, Youth and Sports of the Czech Republic, Hungarian National Research, Development and Innovation Office, New National Excellency Programme of the Hungarian Ministry of Human Capacities, Department of Atomic Energy and Department of Science and Technology of the Government of India, the National Science Centre of Poland, the Ministry  of Science, Education and Sports of the Republic of Croatia, RosAtom of Russia and German Bundesministerium fur Bildung, Wissenschaft, Forschung and Technologie (BMBF), Helmholtz Association, Ministry of Education, Culture, Sports, Science, and Technology (MEXT) and Japan Society for the Promotion of Science (JSPS).

%%%%%%%%%%%%%%%%%%%%%%%%%%%%%%%%%%%%%
%       bibliography
%%%%%%%%%%%%%%%%%%%%%%%%%%%%%%%%%%%%%

\end{document}